\documentclass[12pt,aps,prd,preprint,tightenlines,superscriptaddress,showpacs,nofootinbib]
{revtex4-1}
\pdfoutput=1
\usepackage{amsmath,amssymb,amsthm,amsfonts}
\usepackage{slashed} 
\usepackage{graphicx}
\usepackage{caption}
\usepackage{subcaption}
\usepackage{dcolumn}
\usepackage{hyperref}      
\usepackage{bm}
\usepackage{epsfig}
\usepackage{epstopdf}
\usepackage{setspace}
\usepackage{color}
\usepackage{xcolor,}
\newcommand{\figref}[1]{Fig.~\ref{fig:#1}}

\newcommand{\tableref}[1]{Table~\ref{table:#1}}

\newcommand{\beq}{\begin{equation}}
\newcommand{\eeq}{\end{equation}}
\newcommand{\beqa}{\begin{eqnarray}}
\newcommand{\eeqa}{\end{eqnarray}}

\begin{document}

\title{$H \rightarrow \tau^+ \tau^- \gamma$ as a Probe of the $\tau$ Magnetic Dipole Moment \vspace*{0.25cm}}

\author{Iftah Galon}
\email{iftachg@uci.edu}
\affiliation{Department of Physics and Astronomy, University of
California, Irvine, CA 92697, USA \vspace*{1cm}}

\author{Arvind Rajaraman}
\email{arajaram@uci.edu }
\affiliation{Department of Physics and Astronomy, University of
California, Irvine, CA 92697, USA \vspace*{1cm}}

\author{Tim M. P. Tait}
\email{ttait@uci.edu}
\affiliation{Department of Physics and Astronomy, University of
California, Irvine, CA 92697, USA \vspace*{1cm}}

\preprint{UCI-HEP-TR-2016-16}
\preprint{MITP/16-104}

\begin{abstract}
Low energy observables involving the Standard Model fermions which are chirality-violating, such as anomalous electromagnetic moments,
necessarily involve an insertion of the Higgs in order to maintain $SU(2) \times U(1)$ gauge invariance.  As the result, the properties of the
Higgs boson measured at the LHC impact our understanding of the associated low-energy quantities.  We illustrate this feature with a discussion
of the electromagnetic moments of the $\tau$-lepton, as probed by the rare decay $H \rightarrow \tau^+ \tau^- \gamma$.  We assess the feasibility
of measuring this decay at the LHC, and show that the current bounds from lower energy measurements imply 
that $13~\rm{TeV}$ running is very likely to improve our understanding of new physics contributing to the anomalous magnetic moment of the tau.
\end{abstract}

\maketitle

\section{Introduction}
The discovery of the Higgs boson by the LHC \cite{Aad:2012tfa,Chatrchyan:2012xdj}
in the $\sqrt{s} =7$ and $8~\rm{TeV}$ datasets represents a key milestone in the ongoing study of the Standard Model (SM).
The fact that the observed boson has features which, at least in the broad brush, match with the SM expectations
confirms that it is the principle agent of electroweak symmetry breaking.  A major focus of the LHC run II is to
establish its properties to high precision to either confirm the SM vision for electroweak breaking or to find the
influence of new physics \cite{Morrissey:2009tf,Chang:2008cw,Curtin:2013fra,Belusca-Maito:2016axk}.

As the agent of the electroweak breaking, the observed Higgs boson is connected to any process for which the chirality of
a SM fermion must flip.  To maintain $SU(2) \times U(1)$ gauge invariance, such processes must contain an insertion
of an electro-weak breaking vacuum expectation value (VEV), and the Higgs boson, as the fluctuations around the electroweak vacuum,
has interactions connected with all such terms.  A particularly interesting class of such observables are the electroweak dipole moments
of the SM fermions, dimension five operators which can be sensitive probes of physics beyond the Standard Model.  In fact, a long-standing mystery surrounds
the magnetic dipole moment of the muon, whose experimental determination defies the best available theoretical
predictions of the SM at more than two sigma \cite{Bennett:2006fi}.

The anomalous magnetic dipole moment of a fermion $\psi$ is usually written in terms of the mass $m_\psi$ and the electromagnetic
coupling $e$ as:
\beq
a_\psi~ \frac{e }{2 m_\psi}~ \bar \psi  \sigma^{\mu\nu}\psi ~ F_{\mu\nu}
\label{eq:general_dipole_lowE}
\eeq
where 
$F_{\mu\nu}$ is the electromagnetic field strength tensor and the real quantity
$a_\psi$ parameterizes the size of the anomalous magnetic moment.  The chiral structure of $\sigma^{\mu \nu}$
demands that one of $\psi$ or $\bar \psi$ be left-chiral (and thus part of an $SU(2)$ doublet), and the other right-chiral (and thus an $SU(2)$ singlet).
In the UV theory, it descends from a pair of dimension six operators combining $\bar \psi_L \sigma^{\mu \nu} \psi_R$
with an additional Higgs doublet (and field strengths for the
$U(1)$ and $SU(2)$ gauge bosons).  This obscured dependence on electroweak symmetry breaking is the origin of the
well known fact that, in a theory where the SM Yukawa interactions account for all of the chiral symmetry breaking,
the anomalous magnetic moment is proportional to $m_\psi$ itself.  It also implies that
$a_\psi$ maps uniquely (in a theory with a single source of
electroweak breaking) to an interaction between the Higgs boson, $\psi \bar \psi$, and a photon.

In this article, we focus on the magnetic dipole moment of the $\tau$ lepton.  As the heaviest of the SM 
leptons\footnote{Quark dipole moments are more subtle, manifesting as properties of
the hadrons into which they are bound.}, the $\tau$
is a natural place to search for new physics, and the fact that neutrino masses imply some kind of physics beyond the Standard
Model may be a further indication that the lepton sector is a good place to look for its influence. Indeed,
if the anomalous magnetic moment of the muon is indeed a manifestation of such
physics, one could hope that even larger relative deviations could be present in the $\tau$ sector.  At the same time, measurements
of the $\tau$ dipole moment are currently relatively mildly constrained, leaving room for large deviations.

Higgs decays thus furnish an opportunity to study the $\tau$ magnetic dipole, through the rare decay,
\beq
h \to \tau^+\tau^- \gamma
\label{eq:3body_decay}
\eeq
which we find is potentially observable during high luminosity running of the LHC.  
As explained above, this process is related through gauge invariance to the
more traditional probes of the $\tau$'s electroweak interactions via
precision measurements of its production and/or decay~\cite{
Silverman:1982ft,
Almeida:1991hq,
delAguila:1991rm,
Samuel:1992fm,
Aeppli:1992tp,
Escribano:1993pq, 
Escribano:1996wp,
GonzalezSprinberg:2000mk,
Bernabeu:2007rr,
Bernabeu:2008ii,
Atag:2010ja,
Peressutti:2012zz,
Hayreter:2013vna,
Fael:2013ij,
Hayreter:2015cia,
Eidelman:2016aih},
and we will see that Higgs decays provide both an opportunity to discover physics beyond the Standard model, or
to provide some of the best constraints on an anomalous contribution to the $\tau$ magnetic moment.

This work is structured as follows.
In Section~\ref{sec:constraints} we discuss the operators which
parameterize new physics contributions to the $\tau$ magnetic moment, and review the existing constraints.
In Section~\ref{sec:htta}, we outline the search strategy and discuss backgrounds, and in Section~\ref{sec:limits}
we discuss the potential for discovery or limits in the case no excess of signal events is found.  We
conclude in Section~\ref{sec:conclusions}.

\section{Dipole Operators of the $\tau$ Lepton: Current Constraints}
\label{sec:constraints}

Anomalous contributions to the electroweak dipole moments of the tau (at low-energy scales) originate from dimension six operators:
\beq
c_1 ~\bar \tau_R \sigma^{\mu\nu}B_{\mu\nu} ~H^\dagger L_3
+ c_2 ~\bar \tau_R \sigma^{\mu\nu}~H^\dagger W_{\mu\nu}  L_3  
+ h.c.
\label{eq:dim6}
\eeq
where $L_3$ is the third family left-handed $SU(2)$ doublet, $H$ is the Higgs doublet, $B_{\mu \nu}$ and 
$W_{\mu \nu}$ are the field strengths for
the hypercharge and $SU(2)$ gauge bosons, and $c_1$ and $c_2$ are complex numbers with units of $(\rm{energy})^{-2}$. 
After electroweak symmetry-breaking, linear combinations of these operators lead to anomalous magnetic
and electric dipole interactions, and analogous terms involving the $Z$ boson.  

Currently, the most stringent constraints on the $\tau$ dipole operators are from LEP2, where the
DELPHI collaboration searched for 
$e^+e^-\to e^+e^-\tau^+\tau^-$ events, 
at various collision energies between $183~\rm{GeV}$ and $208~\rm{GeV}$
with a dataset corresponding to an integrated luminosity of $650~\rm{pb}^{-1}$~\cite{Abdallah:2003xd}.
The results were found to be consistent with the SM expectations, leading to the constraint
\beq
-0.052 < a^\gamma_\tau < 0.013,\quad 95\%~{\rm CL}.
\label{eq:agamma_bound}
\eeq
Similarly, the ALEPH collaboration searched for $e^+e^-\to \tau^+\tau^-$ events on the $Z$-pole
with a dataset corresponding to an integrated luminosity of $155~\rm{pb}^{-1}$~\cite{Heister:2002ik}, and obtained the limit
\beq
a^Z_\tau < 1.14 \times 10^{-3} ,\quad 95\%~{\rm CL}.
\eeq
Since the $Z$ magnetic dipole interaction is much more severely constrained, we choose to focus on $a_\tau^\gamma$
from here on.

The combination related to the magnetic moment is described by,
\beq
{1\over \Lambda^2} \bigg\{ 
v ~\bar \tau \sigma^{\mu\nu}~\tau F_{\mu \nu}
+ h~ \bar \tau \sigma^{\mu\nu}~\tau F_{\mu \nu}  
\bigg\} 
\label{eq:mdm}
\eeq
where $h$ is the Higgs boson and $\Lambda^2$ is
a real parameter with dimensions of $(\rm{energy})^2$.  
The anomalous magnetic moment
$a_\tau^\gamma$ is related to $\Lambda$ and the Higgs VEV $v$ via:
\beqa
a^\gamma_\tau &=& -\frac{4 m_\tau}{e}~ {v~ \over \Lambda^2}.
\label{eq:conversion}
\eeqa
In terms of $\Lambda$, the LEP bound (\ref{eq:agamma_bound}) implies
\beq
|\Lambda| > \begin{array}{l} 333~ \rm{GeV} \\
666~\rm{GeV}  \end{array}
~~~~~~~~~{\rm for:}~~~
\begin{array}{r}  ~\Lambda^2 > 0\\
 \Lambda^2 < 0 \end{array}.
\label{eq:LEP_bounds}
\eeq
The second term of Eq.~(\ref{eq:mdm}) leads to the rare Higgs decay 
$h \to \tau^+\tau^- \gamma$.
Measurements of  this decay therefore translate into measurements
of the dipole moment. 
In fact, the SM contribution to this decay mode has the same chirality structure
as the dipole operator, allowing for constructive interference, and resulting in
a relative enhancement of the new physics contribution compared to the
direct search by LEP.

It is worth mentioning that the imaginary parts of $c_1$ and $c_2$ would also lead to
a CP-violating electric dipole moment (and its $Z$ analogue) for the $\tau$.  While interesting
in their own right~\cite{Zhang:2002zw,Bower:2002zx,
Desch:2003mw,
Harnik:2013aja,
Berge:2008wi,
Berge:2008dr,
Berge:2011ij,
Berge:2013jra,
Hagiwara:2016zqz}, CP symmetry prevents these new physics amplitudes from interfering with the
SM contribution to $h \to \tau^+\tau^- \gamma$, thus leading to greatly decreased
LHC sensitivity (see also~\cite{Chen:2014ona}).  For this reason, we choose here to focus on the magnetic dipole moment for which Higgs
decays are a more sensitive probe.

\section{$ h\to\tau^+ \tau^- \gamma$ at the LHC}
\label{sec:htta}

In this section we estimate the LHC sensitivity reach to the $\tau$ magnetic dipole moment through
the decay $h\to\tau^+ \tau^- \gamma$.  This is a challenging signal to reconstruct at a hadron collider for
a number of reasons.  First, $\tau$ leptons decay promptly, producing missing momentum along with either
a charged lepton or a handful of hadrons.  Events containing more than one of them can at best be
incompletely reconstructed.  Decays involving neutral pions also contain energetic photons, which can
potentially fake the additional $\gamma$ which distinguishes our rare decay mode from 
background associated with $h \to \tau^+ \tau^-$.
Photons themselves receive non-perturbative contributions to their production from QCD, which
are not very well understood.  Minimizing these uncontrolled contributions to photon production
typically  requires tight isolation cuts, which can be inefficient for events with dense angular population of energy in the detector.
In addition,
jets can fake both taus and photons at a rate which is intimately tied to the detector response, which is beyond our
ability to reliably estimate.  To work around these difficulties, we base our $h\to\tau^+ \tau^- \gamma$ on the
existing CMS search for $h\to\tau^+ \tau^-$~\cite{Chatrchyan:2014nva},
allowing us to draw from its detailed background study.
We augment this search 
by requiring an additional energetic final state photon
that along with the $\tau^+\tau^-$ system reconstructs the Higgs.

\subsection{Signal Selection}

The CMS search analyzes multiple signal categories, based on the
$\tau$-pair decay modes, the decay products transverse momentum ($p_T$) spectrum,
and $N_j$, the number of high $p_T$ jets in the event.  It is inclusive with regard to photons.
The background estimates in each category are useful
in deducing the dominant background components
when an additional photon is required. 
Moreover, the results for the $N_j + 1$
categories can be used in estimating
the backgrounds from jets faking photons
by scaling the yields with the appropriate fake factor.

We focus on 
categories which have one
hadronic tau ($\tau_h$), and one leptonic tau decaying either to an $e$ or a $\mu$ ($\tau_e$ and $\tau_\mu$, respectively).
This is motivated by the fact that the efficiency of the $\tau$-pair reconstruction method
is larger for the $\tau_h\tau_\ell$ modes than 
for purely leptonic modes which have more neutrinos,
as well as for purely hadronic modes which suffer from uncertainties
related to $\tau$-tagged jets \cite{Chatrchyan:2014nva}. 

We select the ``low-$p_T$'' categories, 
for which the selection criteria for the $\tau$ candidates
are close to the trigger threshold.
This category is particularly sensitive to the $h\to\tau^+ \tau^- \gamma$ decay because:
the additional photon in the Higgs decay implies that
the $\tau^+ \tau^-$ pairs are on average
less energetic than in the two-body decay;
we expect our additional selection requirements
to substantially reduce backgrounds, and
low-threshold signal categories are likely to result
in a larger signal sample to start with.
We further select the $N_j=0$ categories because a Higgs produced recoiling
against additional jets has its
acceptance reduced because
its boosted decay products 
become more collimated,
and tend to fail the isolation criteria more often.

CMS also employs a transverse mass cut, $m_T < 30~\rm{GeV}$, which
substantially reduces $W$-boson related backgrounds.
The final selection cuts for the two search categories ($\tau_h \tau_e$ and $\tau_h \tau_\mu$) are 
shown in \tableref{cut_list_baseline}. For completeness,
we also include the additional cuts of our analysis
which are discussed in subsequent sections.

\begin{table}[t!]
\begin{tabular}{|c|c|c|}
\hline\hline

~~~Decay Mode~~~ & 
~~~~~~~~~~~~CMS Cuts~~~~~~~~~~~~~ & 
~~~~~~~~~~~~Additional Cuts~~~~~~~~~~~~~
\\\hline
& 
$p_T^e > 24, \quad~ |\eta^e| < 2.1$
&
$p_T^{\gamma} > 30~\rm{GeV}$, \quad~$|\eta^\gamma| < 2.5$
\\ 
$\tau_h \tau_e$ 
& ~~$p_T^{\tau_h} > 30, ~\quad |\eta^{\tau_h}| < 2.4$
&
$p_T^{\tau_h} < 45~\rm{GeV}$
\\
&
$m_T < 30~\rm{GeV}$
&
$m_{\tau\tau} < 60~\rm{GeV}$
\\\hline
&
$p_T^\mu > 20, \quad~ |\eta^\mu| < 2.1$
&$p_T^{\gamma} > 30~\rm{GeV}$, \quad~$|\eta^\gamma| < 2.5$
\\ 
$\tau_h \tau_\mu$ 
& ~~$p_T^{\tau_h} > 30, ~\quad |\eta^{\tau_h}| < 2.4$
&
$p_T^{\tau_h} < 45~\rm{GeV}$
\\
&
$m_T < 30~\rm{GeV}$
&
$m_{\tau\tau} < 60~\rm{GeV}$
\\\hline
\hline
\end{tabular}
\caption{Baseline cuts for the signal categories (middle column, from \cite{Chatrchyan:2014nva})
and additional cuts (right column) for search categories with $\tau_h \tau_e$ and $\tau_h \tau_\mu$ tau decays. }
\label{table:cut_list_baseline}
\end{table} 

It is worth mentioning that one could also consider different production topologies for the Higgs in searching for
$h\to\tau^+ \tau^- \gamma$, such as the vector boson + $h$ (VH)~\cite{Aad:2015zrx} or vector boson fusion (VBF)~\cite{Chatrchyan:2014nva}
production modes, which show good sensitivity to $h \to \tau^+ \tau^-$.  These modes yield a richer final state, and thus are somewhat
more fragile with respect to tight photon and tau isolation criteria.
While it would be worthwhile to pursue them as part of a multi-channel analysis of $h\to\tau^+ \tau^- \gamma$, we leave their
detailed investigation for future work.

\subsection{Backgrounds}

A detailed study of the backgrounds contributing to the search for $h \to \tau^+ \tau^-$ is presented in the CMS study, Ref.~\cite{Chatrchyan:2014nva}.
In this section, we use these results to infer the most important backgrounds for $h \to \tau^+ \tau^- \gamma$ in the $\tau_h \tau_\ell$, $N_j=0$ topology.
The dominant background for $h \to \tau^+ \tau^-$ arises from production of a $Z$ boson which subsequently decays into $\tau^+ \tau^-$.
There are also much smaller contributions from electroweak and QCD processes.  We further require an additional energetic isolated photon to
be present, which can either correspond to real electromagnetic radiation or a jet which is misidentified as a photon.

The contributions to the real photon backgrounds can be roughly estimated from the $N_j=0$ backgrounds studied in
\cite{Chatrchyan:2014nva} scaled\footnote{Note that the required stringent isolation cuts control large collinear logarithms.}
by  $\sim \alpha$.  This scaling suggests that of the backgrounds considered, only the $Z \to \tau^+ \tau^-$ is large enough to
be relevant.  Thus, the real photon background can be approximated as coming entirely from $Z+\gamma$ diboson production.

The size of the background from jets misidentified as a photon can be estimated based on the $N_j=1$ analysis of~\cite{Chatrchyan:2014nva}
scaled by the fake rate of $f_{j\to\gamma} \sim 10\%$~\cite{CMS:2015loa}.  
To make a fair comparison, we apply a cut of $p_T > 30~\rm{GeV}$ and $|\eta^\gamma| \leq 2.5$ on signal photons in order
to match the jet cut in the $N_j =1$ categories.
Once again, the only relevant background after applying the fake rate is $Z+$jet production, where the
$Z$ decays into taus and the jet fakes a photon.

The backgrounds can be reduced by applying further cuts on top of the CMS analysis.
First, we require $p_T^{\tau_h}<45~\rm{GeV}$ (see also~\cite{CMS:2015mca}), which reduces contributions from $Z$ (and $W$) boson decays,
while having negligible effects on the signal yield.
Second, we impose a cut on the $\tau^+ \tau^-$ invariant mass, $M_{\tau\tau}< 60~\rm{GeV}$ in order to reduce 
the $Z$-background which is narrowly centered around
$m_Z \sim 91~\rm{GeV}$.
In a realistic setting, the $M_{\tau\tau}$ distribution is smeared out by the imperfect
$\tau$ reconstruction; nonetheless we will see below that
an upper cut $M_{\tau \tau} < 60-75~\rm{GeV}$ 
removes most of the $Z$ background, while
still preserving a large part of the signal.

\subsection{Monte Carlo Simulation}

To assess future LHC sensitivity to the dipole operator from searches for the $h \to \tau_\ell \tau_h \gamma$ decay mode, we perform a Monte Carlo
simulation of the signal and $Z+$$\gamma$ and $Z+$jet backgrounds described above.  All three processes are simulated
in MadGraph5\_aMC@NLO (MG5)~\cite{MG5}, with showering and hadronization by Pythia6~\cite{PYTHIA6,pythia-pgs}, and jet matching
under the CKKW prescription~\cite{Catani:2001cc,Krauss:2002up,Alwall:2007fs}.  Tau decays are handled at the generator level~\cite{Hagiwara:2012vz}
(as opposed to by Pythia), as discussed below.  The hard matrix elements are derived from FeynRules implementations~\cite{Degrande:2011ua,Alloul:2013bka},
supplemented by higgs effective vertices with gluons and photons and the tau dipole operators~\cite{Alloul:2013naa}. 
For the $Z+$jet background, we include a K-factor of $K_{Zj}\sim 1.5$ representing
the enhancement of the cross-section from higher order QCD corrections \cite{Ridder:2015dxa}.

The detector reconstruction is simulated by the Delphes 3.3.0~\cite{Delphes} detector emulator
with parameters from its default CMS card.
At the detector level,
anti-$k_T$ jets~\cite{AntiKt} are reconstructed with FastJet~\cite{Cacciari:2011ma}.
Electrons, muons, and photons are required to be isolated within a cone of size $\Delta R = 0.5$, where an object is considered to be isolated  if 
the ratio of the sum of $p_T$ depositions within the cone around it to its own $p_T$ is smaller than $0.1$.
Photon isolation, is applied in MG5 using the built-in Frixione prescription~\cite{Frixione:1998jh} with parameters 
$\Delta R=0.4$ and $p_T^{\gamma~min} = 10~\rm{GeV}$.  We decay the taus at the generator level in order to
include the pions from their decay in this isolation cut.

The tau pair kinematics are reconstructed using the public stand-alone version of the package, 
SVFitStandAlone~\cite{Bianchini:2014vza}, employing a 
likelihood based method~\cite{Elagin:2010aw,Chatrchyan:2014nva,SVFitStandAlone}
which improves on the collinear approximation~\cite{Ellis:1987xu}.
We calibrate this package with the Delphes-level covariance matrix of the transverse missing-energy two-vector $\vec {\slashed E}_T$
using a Monte Carlo sample of $Z+$jet in the CMS fiducial region. 
By comparing the ``truth''-level missing energy
information ( neutrinos ) with the reconstructed $\vec {\slashed E}_T$,
we deduce the covariance matrix parameters.
We find that SVFitStandAlone reconstructs masses for the $h$ and $Z$ that are systematically somewhat higher 
than their pole masses.  As a result, the Higgs peak is smeared into the range $120-140~\rm{GeV}$.
We thus define 
$120~\rm{GeV} \le M_{\tau\tau\gamma} < 140~\rm{GeV}$ as the signal region of our analysis.

We note that Ref.~\cite{Chatrchyan:2014nva} also imposes a cut on the energy deposited near the hadronically decaying tau.
This cut is not possible to implement via our work-flow, and thus is neglected.  However, because of the strong isolation
cuts already imposed on the other hard final state objects in the event, we expect this omission has little effect on our
final conclusions.

\section{Results and Projected Limits}
\label{sec:limits}

\begin{figure}[t!]
\centering
\includegraphics[width=0.47\textwidth]{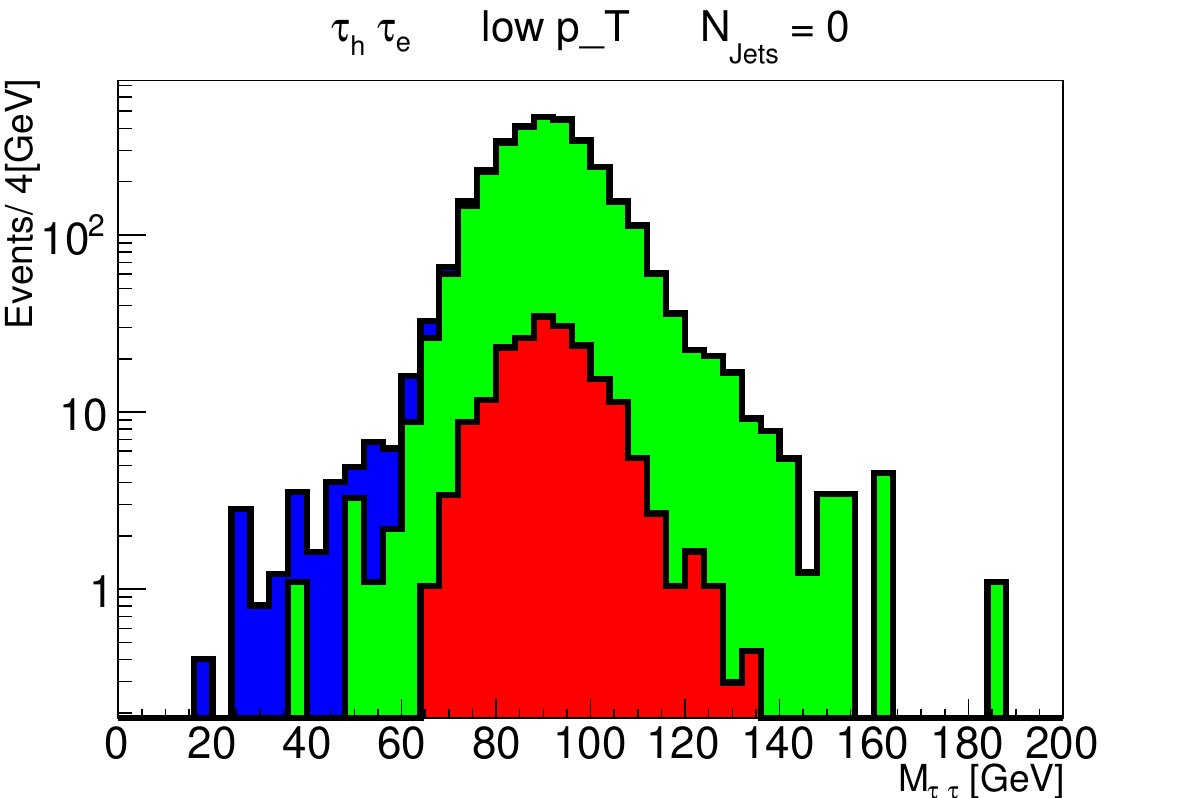}
~
\includegraphics[width=0.47\textwidth]{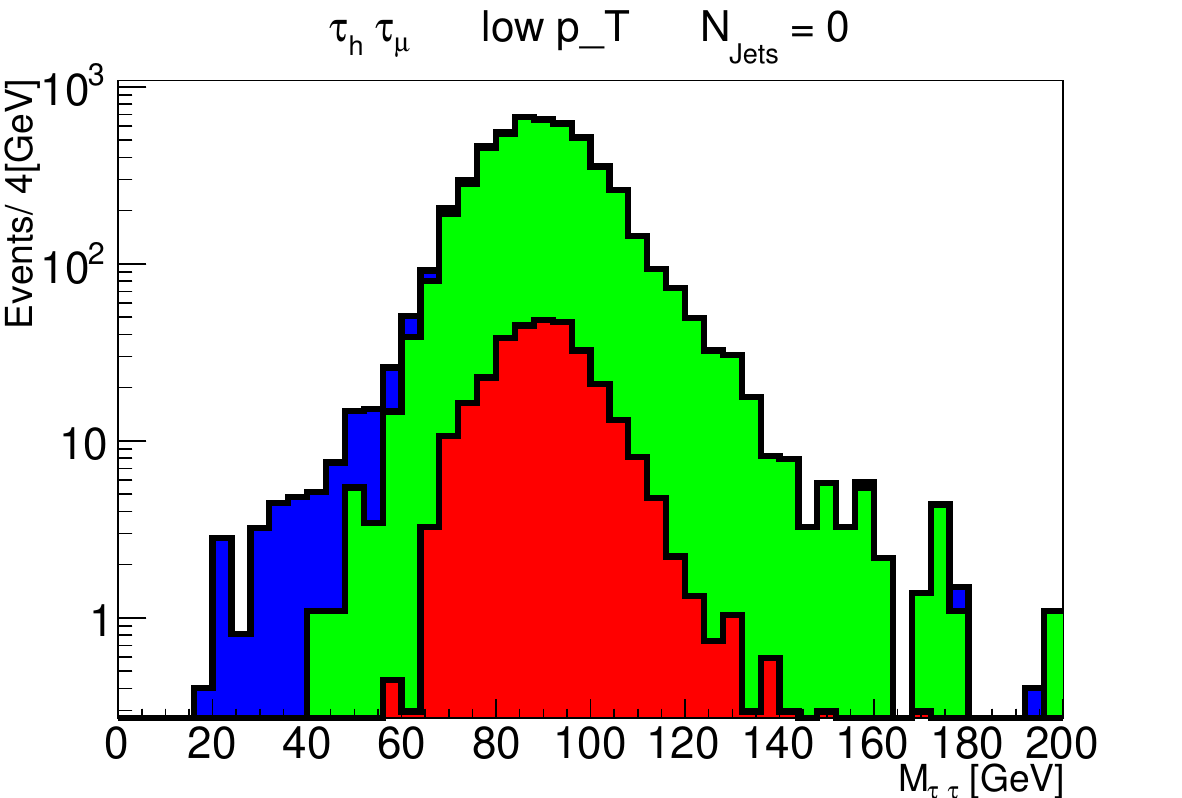}
\\[0.5cm]
\includegraphics[width=0.47\textwidth]{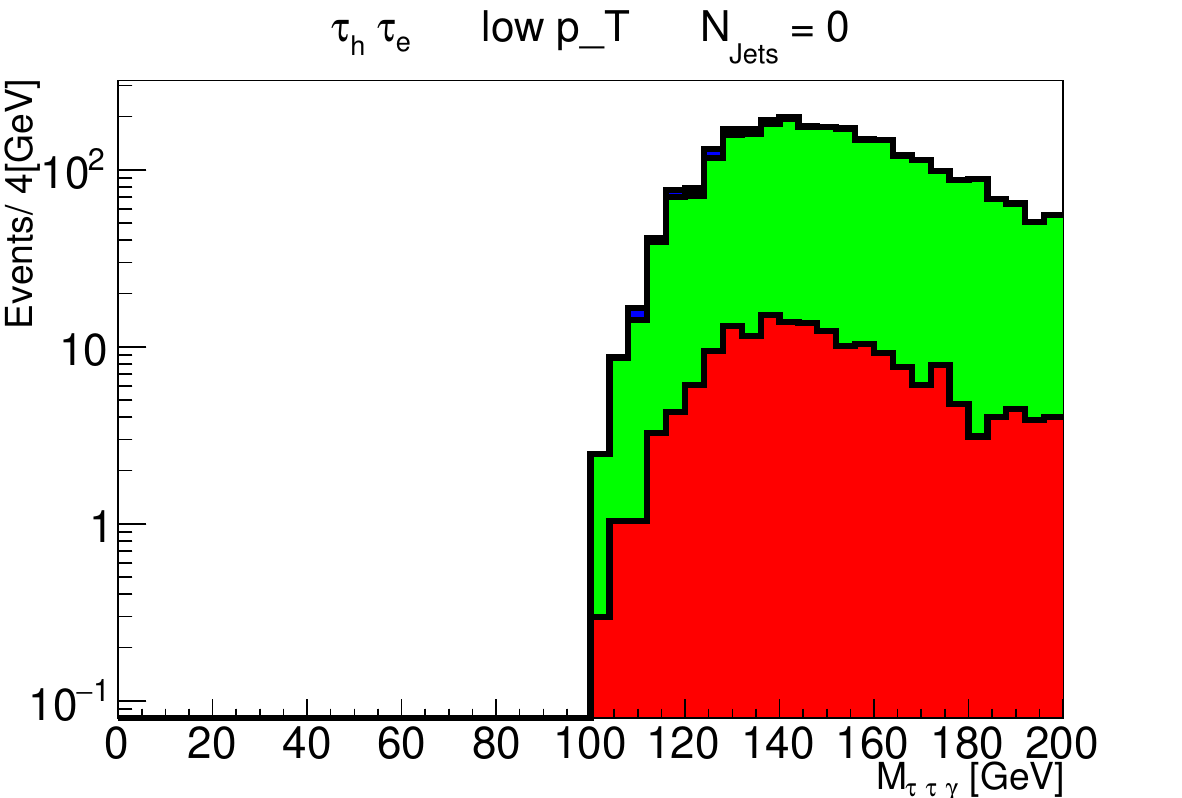}
~
\includegraphics[width=0.47\textwidth]{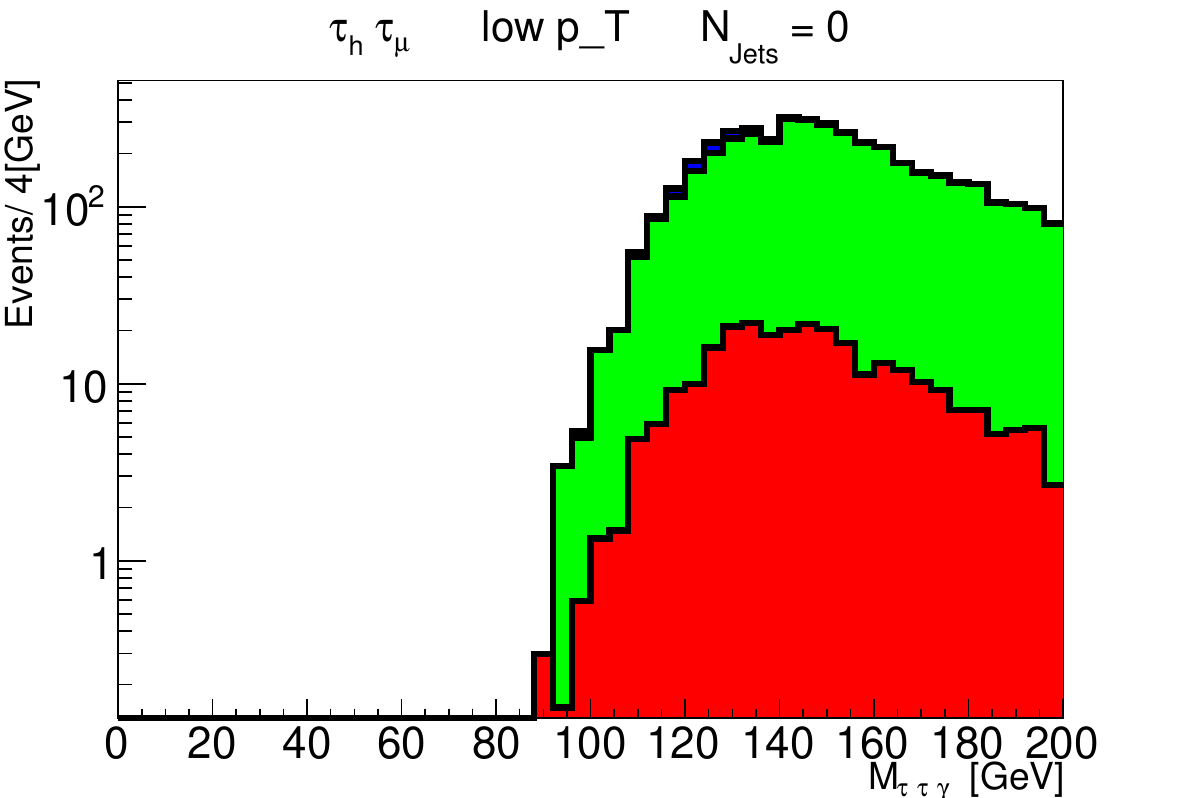}

\caption{
The $M_{\tau\tau}$ (upper plots) and $M_{\tau\tau\gamma}$ (lower plots) distributions
for the two signal categories $\tau_h\tau_e$ (left) and $\tau_h\tau_\mu$ (right)
at $\sqrt{s}=13~\rm{TeV}$ and ${\cal L}_{\rm int} = 300~\rm{fb}^{-1}$,
before applying the $M_{\tau\tau} < 60~\rm{GeV}$ cut.
Contributions from the signal $gg\to h\to \tau\bar\tau\gamma$ (blue),
$pp\to Z+\gamma$ (red), and $pp\to Z+j(\gamma)$ (green) are indicated.
}
\label{fig:13TeV_histo_stacked_no_mtautaucut}
\end{figure}

\begin{figure}[t!]
\centering
\begin{subfigure}[b]{0.45\textwidth}
\includegraphics[width=\textwidth]{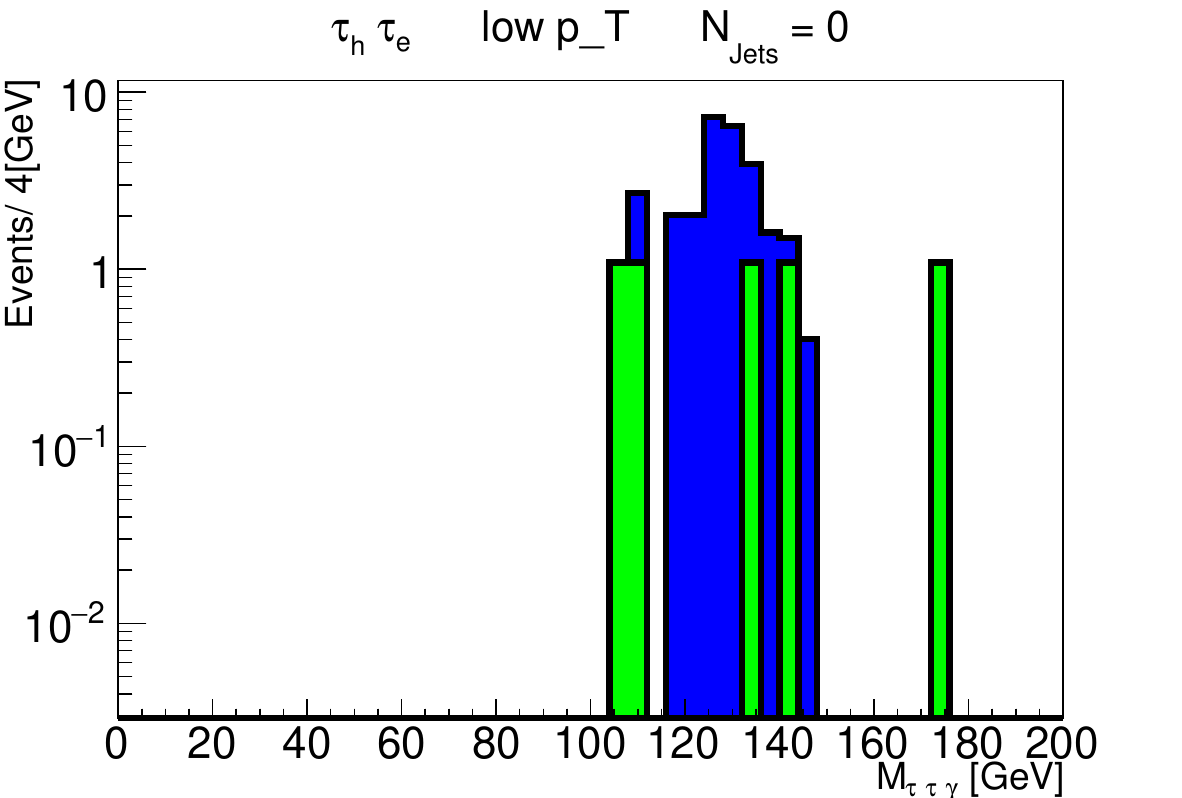}
\end{subfigure}
~
\begin{subfigure}[b]{0.45\textwidth}
\includegraphics[width=\textwidth]{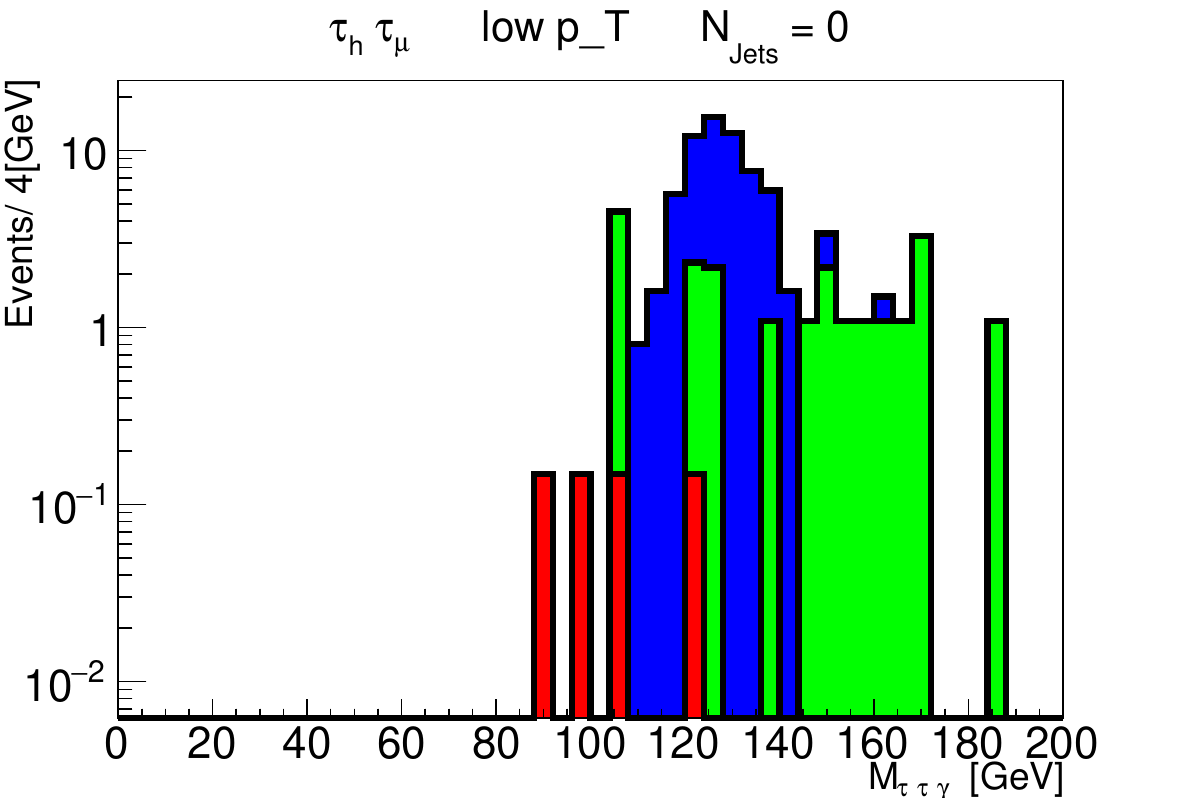}
\end{subfigure}

\caption{The $M_{\tau\tau\gamma}$ distributions of the two signal categories, $\tau_h\tau_e$ (left) and $\tau_h\tau_\mu$ (right),
after applying the $M_{\tau\tau} < 60~\rm{GeV}$ cut, with color coding as in \figref{13TeV_histo_stacked_no_mtautaucut}.
 }
\label{fig:13TeV_histo_stacked_mtautaucut}
\end{figure}

We analyze the reach of the LHC running at $\sqrt{s} = 13~\rm{TeV}$, with the reconstruction strategy outlined above.
In \figref{13TeV_histo_stacked_no_mtautaucut}
we show the expected $M_{\tau \tau}$ and $M_{\tau\tau\gamma}$ distributions for the
signal (with  $\Lambda = 343~\rm{GeV}$, $\Lambda^2 >0$) and $Z+\gamma$ and $Z+j$ background processes,
in the $\tau_h \tau_e$ and $\tau_h \tau_\mu$ topologies,
for an integrated luminosity of ${\cal L}_{\rm int} = 300~\rm{fb}^{-1}$.
All analysis cuts with the exception of the $M_{\tau\tau} < 60~\rm{GeV}$ cut are applied.  Evident from the upper plots
is the motivation for the $M_{\tau\tau} < 60~\rm{GeV}$ cut to separate the signal from the backgrounds.
The $M_{\tau\tau\gamma}$ distributions after this cut are presented in 
\figref{13TeV_histo_stacked_mtautaucut}, and demonstrate its dramatic efficacy, with only a
handful of background events left in the $120~\rm{GeV} \le M_{\tau\tau\gamma} < 140~\rm{GeV}$ signal region.

\begin{figure}[t!]
\centering
\includegraphics[width=0.475\textwidth]{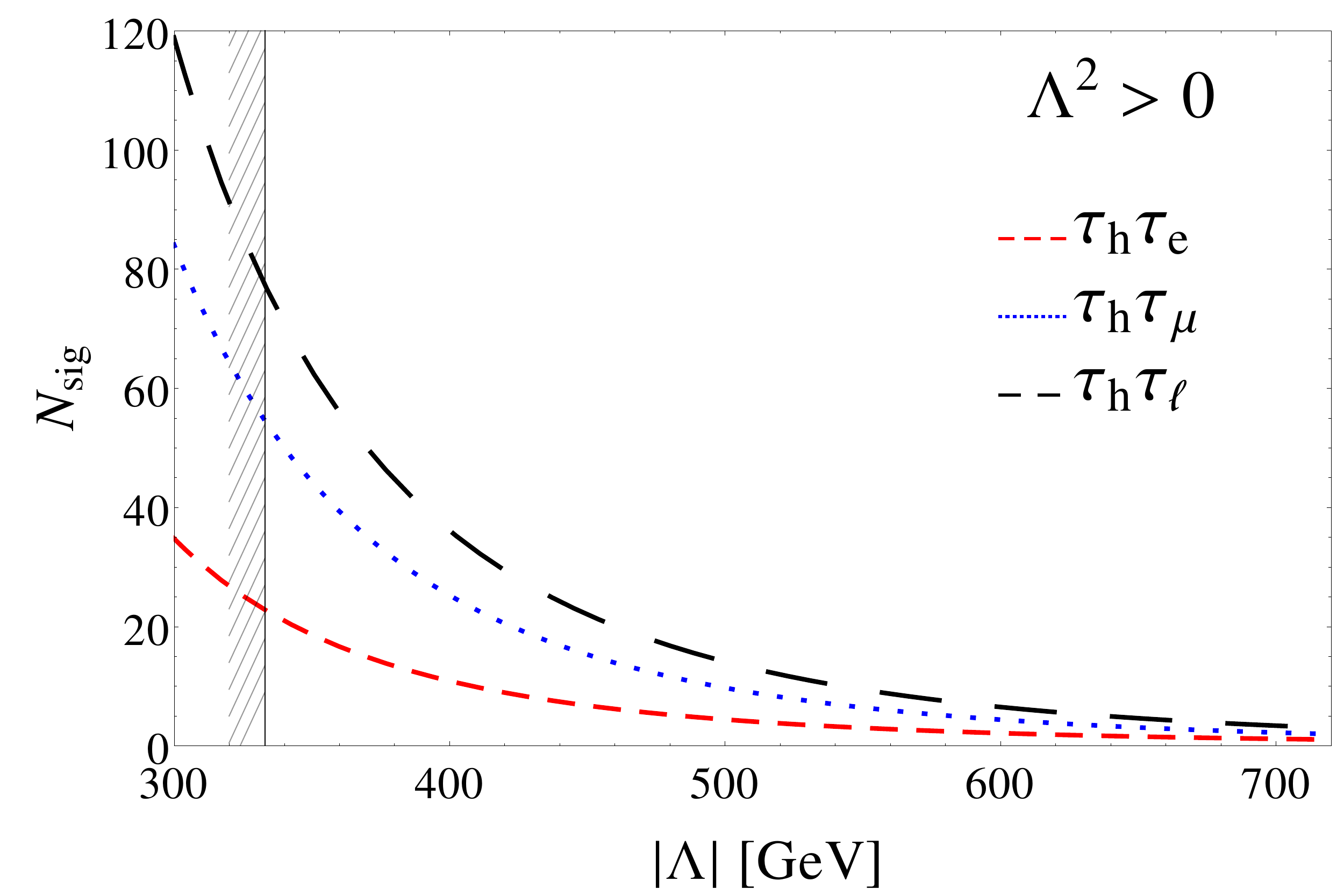}
~
\includegraphics[width=0.475\textwidth]{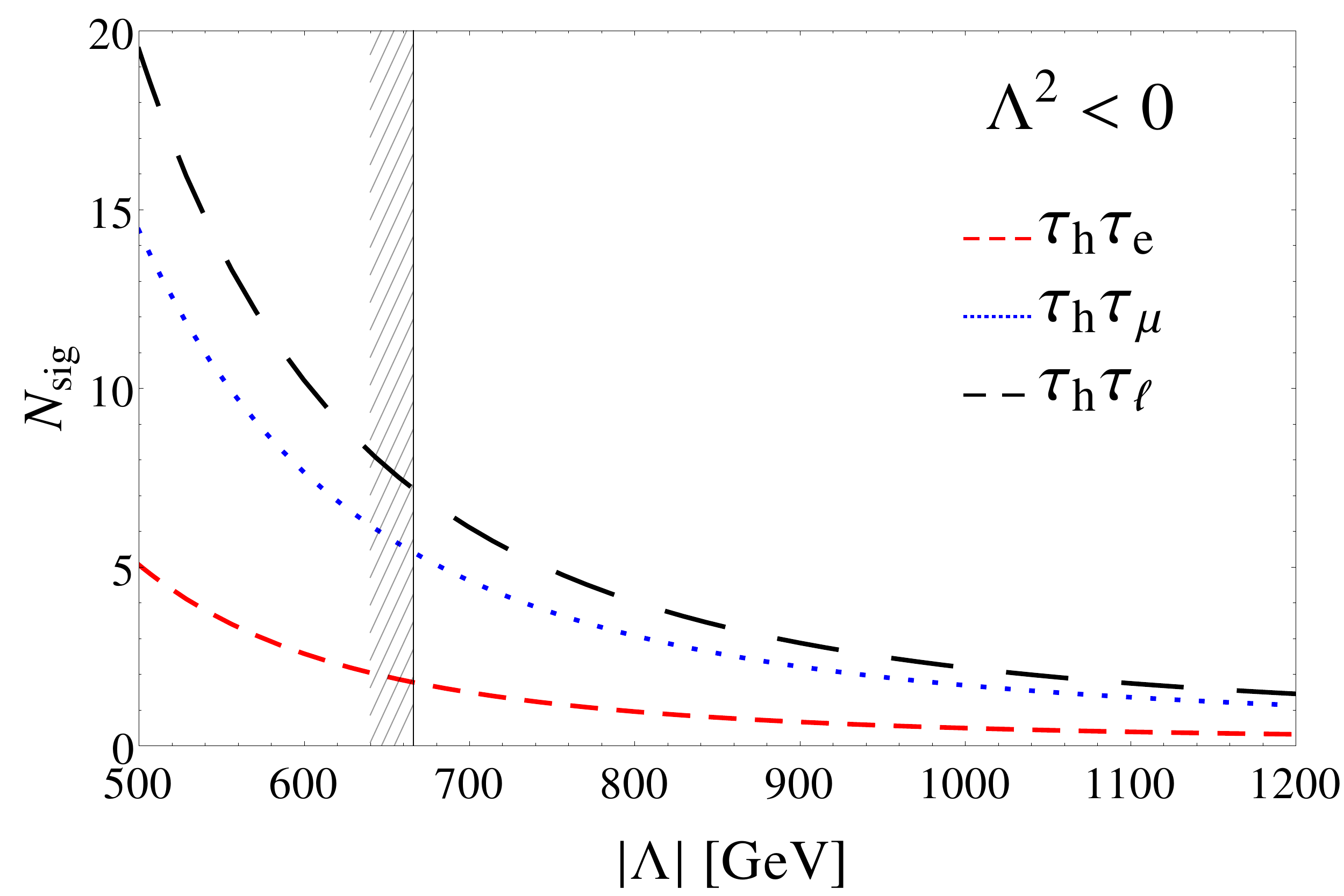}
\caption{The expected number of signal events, $N_{\rm{sig}}$, as a function of $\Lambda$ in the two 
signal categories, $\tau_h\tau_e$ (red), $\tau_h\tau_\mu$ (blue),
and their combination $\tau_h\tau_\ell$ (black),
for $\Lambda^2> 0$ (left plot) and $\Lambda^2 <0$ (right plot),
for ${\cal L}_{\rm int} = 300~\rm{fb}^{-1}$.
The LEP exclusion limits are indicated by the hatched vertical lines.
}
\label{fig:NsigVSLambda}
\end{figure}

To estimate the eventual sensitivity of the high luminosity LHC to new physics in the
tau magnetic dipole, we write the amplitude for the signal process as
\beq
{\cal M}_{\rm sig} = {\cal M}_{\rm{SM}} + \frac{1}{\Lambda^2} {\cal M}_{\rm{NP}}
\eeq
with the $\Lambda$ dependence explicitly factored out.
The yield of signal events (for ${\cal L}_{\rm int} = 300~\rm{fb}^{-1}$) after cuts is:
\beqa
N_{\rm sig} &=& 
N_{\rm{SM}} + 
\frac{2}{\Lambda^2}
\hat{N}_{\rm INT}
+ \frac{1}{\Lambda^4}
\hat{N}_{\rm NP}.
\label{eq:Nsig}
\eeqa
The sizes of these three coefficients, after all analysis cuts, are shown in Table~\ref{table:coeff_sol}.
We present $N_{\rm sig}$ as a function of $\Lambda$ for both signs of $\Lambda^2$
in Fig.~\ref{fig:NsigVSLambda}, including the $\tau_h\tau_e$ and $\tau_h\tau_\mu$ channels,
and also the combined number of events in both signal categories, $\tau_h\tau_\ell$.

\begin{table}[t!]
\begin{tabular}{|c|c|c|c|c|c|}
\hline\hline
~~~Signal Region~~~ & ~~~~~$N_{\rm{SM}}$~~~~~ & ~~~~~~~~$\hat{N}_{\rm INT}$~~~~~~~~ & ~~~~~~~~$\hat{N}_{\rm NP}$~~~~~~~~
\\\hline
$ \tau_h\tau_e$ 
& $0.124$ 
& ~~$-8.19 \times 10^4$~GeV$^2$ ~~
& ~~$2.88\times 10^{11}$~GeV$^4$~~
\\ \hline
$ \tau_h\tau_\mu$ 
& $0.371$ 
& ~~$-5.88\times 10^{5}$~GeV$^2$~~
& ~~$7.31\times 10^{11}$~GeV$^4$~~
\\\hline
\hline
\end{tabular}
\caption{Sizes of the three coefficients, $N_{\rm{SM}}$, $\hat{N}_{\rm INT}$, and $\hat{N}_{\rm NP}$, corresponding to
${\cal L}_{\rm int} = 300~\rm{fb}^{-1}$.
}
\label{table:coeff_sol}
\end{table} 

Under the Assumption that no signal is observed, and the mean number of background events, 6.7, is obtained, one may place a lower
limit on the new physics scale $\Lambda$.  The $95\%$~CL bound on $\Lambda$ in that case would be given by:
\beq
|\Lambda| > \begin{array}{l} 634~ \rm{GeV} \\
739~\rm{GeV}  \end{array}
~~~~~~~~~{\rm for:}~~~
\begin{array}{r}  ~\Lambda^2 > 0\\
 \Lambda^2 < 0 \end{array}.
\label{eq:LEP_bounds}
\eeq
which translates into a projected limit on the anomalous moment of
\beq
-0.0144 < a^\gamma_\tau < 0.0106. \quad (95\%~{\rm CL}),
\eeq
approximately a factor of two improvement on $\Lambda$ for $\Lambda^2 > 0$ or $10\%$ for $\Lambda^2 < 0$.

\section{Conclusions and Outlook}
\label{sec:conclusions}

As measurements of the Higgs boson become more sophisticated, we move into a regime where it becomes a tool 
in its own right to search for new physics.  In particular, low energy observables involving a chiral flip of the SM
fermions necessarily invoke electroweak symmetry-breaking, and thus imply a modification of the properties of the Higgs.
In this work, we have examined the possibility that one can place bounds on the electromagnetic moments of the $\tau$
by searching for the rare decay of the Higgs into $\tau^+ \tau^- \gamma$.  
Given the longstanding discrepancy between measurements
of the muon's magnetic moment and SM predictions, one could hope that $a^\gamma_\tau$ might also be a likely
target for which to search for manifestations of new physics.

We find the promising result that the LHC with a large data set should be sensitive to modifications of the $\tau$
magnetic dipole moment beyond the current bounds extracted from LEP -- by about a factor of
two on the new physics scale if $\Lambda^2 > 0$.  Given these promising results, it would be worthwhile
to follow up with a study based on more realistic detector simulations and including effects beyond our ability to reliably simulate,
such as pile-up.  We hope that our study will motivate the experimental collaborations to carry out this effort.

Another interesting direction for the future  would be to study the prospects at a future $e^+ e^-$ collider such as the ILC or
a future circular collider~\cite{Ozguven:2016rst}. While the rate for $hZ$ production at such colliders is considerably smaller than the inclusive Higgs
production at the LHC, new physics saturating the LEP bound nonetheless allows for a handful of events, and the prospects
depend sensitively on the ability to efficiently reconstruct the signal events and reject backgrounds.  

The discovery of the Higgs is a triumph of run I of the LHC.  We look forward to run II and beyond
to follow up with precision measurements that reveal the deep secrets that reflect its character.

\section*{Acknowledgments}

The authors are supported in part by National Science Foundation 
grants PHY-1316792 and PHY-1620638.
We thank Enrique Kajomovitz, Olivier Mattelaer, Andrew Nelson, Yoram Rozen, Yael Shadmi, Yotam Soreq, Philip Tanedo 
and Scott Thomas for useful discussions;
Christian Veelken for his assistance with setting up the tau reconstruction platform and for useful discussions; and
David Cohen for assistance in grid-based computing.
IG and TMPT are grateful to the Mainz Institute for Theoretical Physics 
for its hospitality and its partial support during the completion of this work.

\bibliography{htautauaBIB}
\end{document}